\documentclass[preprint]{vgtc}               

\graphicspath{{figures/}{pictures/}{images/}{./}} 

\usepackage{times}                     

\usepackage{tabu}                      
\usepackage{booktabs}                  
\usepackage{lipsum}                    
\usepackage{mwe}                       
\usepackage{amsmath}
\usepackage{mathptmx}                  
\usepackage{xcolor}
\usepackage{microtype}

\onlineid{0}

\vgtccategory{Research}

\vgtcinsertpkg

\newcommand{\appname}{{\texttt{VIS-ReAct}}}
\newcommand{\appnamep}[1][VIS-Act]{{\texttt{#1}}}

\title{
Agentic Reasoning and Refinement through Semantic Interaction
}

\author{
Xuxin Tang\thanks{e-mail: xuxintang@vt.edu}\\ %
    \scriptsize Virginia Tech %
\and Rehema Abulikemu\thanks{e-mail: rexime@vt.edu} \\
    \scriptsize Virginia Tech
\and Eric Krokos\\%
    \scriptsize US Department of Defense
\and Kirsten Whitley\\ %
    \scriptsize US Department of Defense
\and Xuan Wang\thanks{e-mail: xuanw@vt.edu}\\
    \scriptsize Virginia Tech
\and Chris North\thanks{e-mail: north@cs.vt.edu}\\%
    \scriptsize Virginia Tech %
}

\teaser{
  \centering
  \includegraphics[width=\linewidth]{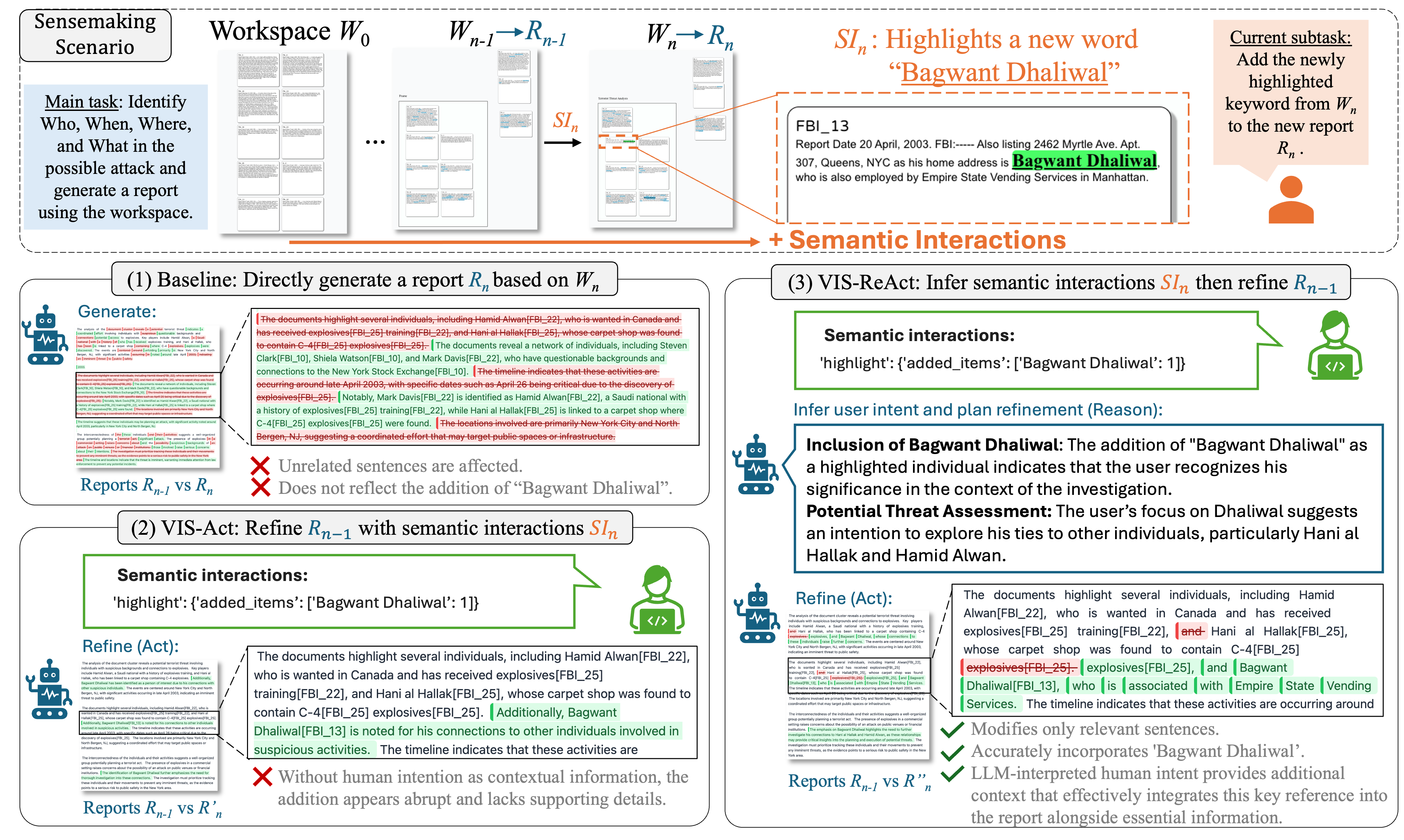}
  \caption{In this sensemaking scenario, each incremental user-generated workspace $W_i$ ($i>0$) corresponds to an LLM-generated report. To evaluate how to incorporate newly-added semantic interactions $SI_n$ in refining reports, we examine a case where a user highlights a suspicious name in $W_n$. We compare three prompting methods: (1) direct summarization (Baseline); (2) our proposed method \appnamep{}, which refines reports based on $SI_n$; and (3) \appname{}, an augmented refinement method with LLM-analyzed $SI_n$. We only presented the main different part in the prompt while omitting shared input. The unified visualization displays deletions from the previous report version in red and additions in green, clearly demonstrating \appname{}'s superior performance in automatic report refinement.
  }
  \label{fig:teaser}
}

\abstract{
    Sensemaking report writing often requires multiple refinements in the iterative process. While Large Language Models (LLMs) have shown promise in generating initial reports based on human visual workspace representations, they struggle to precisely incorporate sequential semantic interactions during the refinement process. We introduce \appname{}, a framework that reasons about newly-added semantic interactions in visual workspaces to steer the LLM for report refinement. 
    \appname{} is a two-agent framework: a primary LLM analysis agent interprets new semantic interactions to infer user intentions and generate refinement planning, followed by an LLM refinement agent that updates reports accordingly. Through case study, \appname{} outperforms baseline and \appnamep{} (without LLM analysis) on targeted refinement, semantic fidelity, and transparent inference. Results demonstrate that \appname{} better handles various interaction types and granularities while enhancing the transparency of human-LLM collaboration.
}


\begin{document}

\maketitle
\section{Introduction}
\begin{figure*}[htb]
 \centering
 \includegraphics[width=0.8\textwidth]{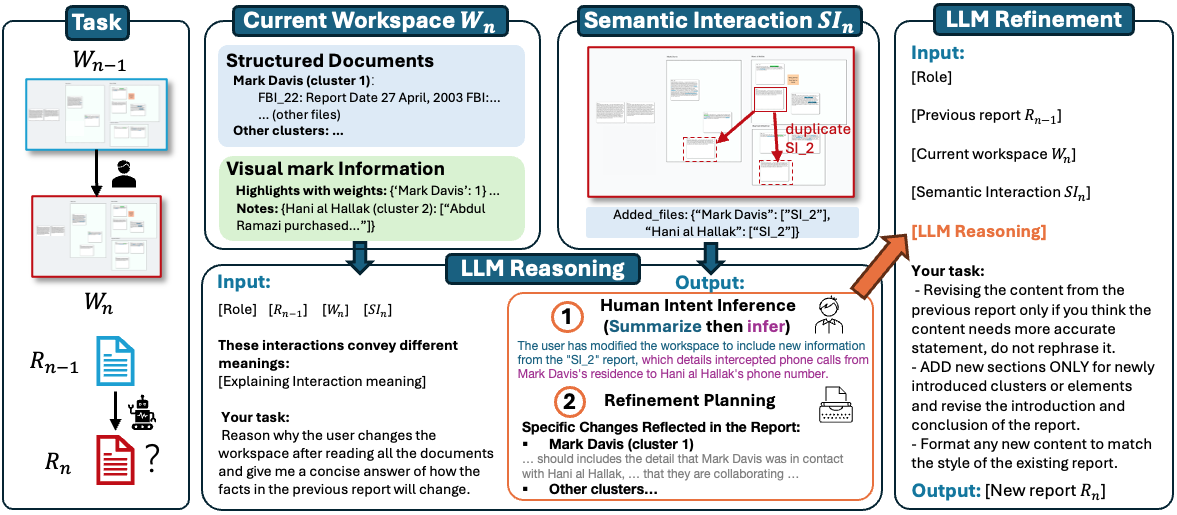}
 \caption{
The \appname{} framework automatically generates refinement reports from semantic interactions through a systematic four-step process: (1) converting the \textbf{current workspace} to text format, (2) extracting \textbf{semantic interaction} data, followed by a two-phase refinement comprising \textbf{LLM analysis} and \textbf{LLM refinement}.
 }
 \label{fig:process}
\end{figure*}
As the final step of multi-document sensemaking, report writing requires analysts to summarize their hypothesis with extracted and connected information in previous cognitive stages \cite{pirolli2005sensemaking}.
Large language models (LLMs) have been introduced in both sensemaking \cite{kang2023synergi,suh2023sensecape,gero2024supporting,tang2025respire} and writing assistants \cite{zhang2023visar,buschek2024collage,lee2024design,tang2024steering,masson2024visual}. 
While sensemaking is an iterative process, after the first draft generated by the LLMs, users may have the demand to refine the draft. How users' modifications can be precisely captured and reflected in the newly-generated text is significant for the human-AI collaboration.

Direct-manipulated visual workspaces like Space to Think \cite{andrews2010space} are utilized to enable users to utilize the spatial organization of documents and visual marks to leave their intermediate insights and externalize cognitive models. ReSPIRE \cite{tang2025respire} implemented the workspace-steered generation proposed by Tang et al. \cite{tang2024steering} by utilizing visual workspaces as the common ground for human-AI sensemaking. In this approach, human feedback—in the form of highlights, notes, and spatial clusters—steers the LLM report generation process. However, due to the randomness of LLM generation, even with fixed temperature, the LLMs will generate different reports in word level for a simple-modified workspace (Fig.~\ref{fig:teaser}(1)). 
The inconsistent modifications across reports create obstacles for users during the refinement process. These discrepancies make it difficult to track corresponding changes, leading to frustration in the human-LLM collaboration workflow.
To address the challenges, we propose the framework \appname{}, 
which achieves targeted sensemaking report refinement through human semantic interactions in visual workspaces. As illustrated in Figure~\ref{fig:teaser}, baseline approach (1) fall short: completely regenerated report using ReSPIRE \cite{tang2025respire} (Figure~\ref{fig:teaser}(1)) affects irrelevant sentences and does not reflect the newly-added highlight keyword. To automate the report refinement without manual prompting, we proposed \appnamep{}, which can targetedly update reports based on semantic interactions (Figure~\ref{fig:teaser}(2)). However, the loss of contextual information makes the added texts abrupt and lacks supporting details. Therefore, we developed advanced \appname{} (Figure~\ref{fig:teaser}(3)) which resolves these issues through a three-step process: it first collects semantic interactions from workspace comparison, then employs an LLM agent to analyze them, infer human intent, and generate contextual refinement planning, and finally uses another LLM agent to refine the reports according to this detailed planning.




Our case study demonstrates that this framework effectively addresses various semantic modifications and granularities in report refinement, improving both targeted updates and content fidelity. The analysis agent enhances human-AI collaboration transparency by providing clear interpretations of user interactions, helping users better track and understand LLM-generated changes.

\section{Related Work}
Sensemaking report generation is a specialized form of abstractive summarization \cite{gupta2019abstractive} with a focus on extracting and connecting specific types of information (clues and evidence) across multiple sources. While most research in abstractive summarization concentrates on generation techniques, relatively few studies address human-in-the-loop refinement processes. ReSPIRE \cite{tang2025respire} incorporates the workspace-steered report generation approach proposed by Tang et al. \cite{tang2024steering}, though it generates completely paraphrased content. Similarly, ConceptEva \cite{zhang2023concepteva} generates summaries supporting both manual editing and LLM paraphrasing of individual elements. LexGenie \cite{Santosh2025LexGenieAG} refines report structure through user-adjustable ranked lists of retrieved information.

Our research addresses a critical question: How can users achieve step-by-step updates and refinements through interactive control mechanisms? Our method addresses incremental formalism \cite{shipman1999formality} by integrating key insights from several research areas: the offloaded spatial cognitive model in Space to Think \cite{andrews2010space}, the intent inference approach of semantic interaction \cite{endert2012semantic}, the user-steered report generation through visual workspaces \cite{tang2024steering}, and the agentic reasoning and action framework of ReAct \cite{yao2023react}. This synthesis creates a more flexible and responsive approach, \appname{}, to human-steered sensemaking report refinement.

\section{\appname{}}
\appname{} enhances ReSPIRE by enabling targeted, context-aware report refinement. ReSPIRE \cite{tang2025respire} supports iterative sensemaking through an interactive visual workspace where users can manipulate documents and visual marks to steer LLMs to generate personalized reports. However, each LLM generation produces an entirely new report, making it challenging to update only relevant portions and enable users to focus on specific changes.

The idea of \appname{}\footnote{Reason and Act framework \cite{yao2023react} for tracking semantic interactions in VISual analytics.} is straightforward, helping users automatically refine and augment the LLM-generated reports through human newly-added semantic interactions $SI_n$ in the visual workspaces. 
Compared to manual prompting, \appnamep{} summarizes $SI_n$ to produce partially refined reports. However, relying solely on $SI_n$ results in lost key contextual information, making refinements vague and superficial. \appname{} analyzes workspace interactions to generate contextual refinement guidance, enabling LLMs to produce targeted and relevant report updates.

As shown in Figure~\ref{fig:process}, consider a scenario during the sensemaking that the user already has a structured workspace $W_{n-1}$ with a LLM-generated report $R_{n-1}$ but wants to refine the report. The framework of \appname{} includes the following steps:



\textbf{Converting the current workspace to text format.} For visual sensemaking, we leverage the ReSPIRE system \cite{tang2025respire} to transform visual workspace into structured text. This conversion produces two key components: structured document representations and visual mark information. While ReSPIRE organizes content using explicit frame-based clustering in a hierarchical cluster-document structure, we also utilize it to capture interaction data—including weighted highlights (with frequency as weights) and notes attached to either clusters or individual documents.

\textbf{Extracting semantic interaction data.}
Research demonstrates that LLMs perform better when provided with the most important data \cite{li2024superfiltering}. We therefore compare previous and current workspaces to extract semantic interactions that meaningfully alter the workspace content. Following established semantic interaction frameworks \cite{endert2012semantic}, our extracted semantic interactions $SI_n$ include cluster creation, deletion, and reorganization. Additionally, we track visual mark modifications, covering the addition, removal, and editing of highlights and notes.

\textbf{LLM analysis.}
As the reasoning component of \appname{}, the LLM analysis module plays a crucial role in examining semantic interactions, inferring user intent, and developing detailed refinement strategies. As illustrated in Figure~\ref{fig:process}, this module processes comprehensive contextual information, including the previously generated report $R_{n-1}$, semantic interaction data $SI_n$, and current workspace state $W_n$. The output comprises two distinct components: human intent inference, which articulates how the LLM interprets user semantic interactions, and refinement planning, which specifies how the subsequent LLM agent will execute targeted refinements.


\textbf{LLM refinement.}
After obtaining the LLM analysis report, LLM refinement is activated with current workspace $W_n$ and previous report $R_{n-1}$. The prompt constrains the LLM to avoid rephrasing, restricts the scope of modifications, and enforces adherence to the specified report format. The analysis and refinement results can be found in the supplemental material.

\section{Case Study}
The visual workspace's flexibility allows users to personally interact with it through semantic interactions \cite{endert2012semantic}, including document reorganization, text highlighting, and annotation. This adaptability lead us to wonder: Can these personal semantic modifications serve as feedback to let LLMs refine the report like humans? 
To address this challenge, we propose that workspace-based report refinement should follow three equally important principles: (\textbf{P1: Targeted Refinement}) refinements should precisely target only the relevant sections \cite{birnholtz2012tracking}; (\textbf{P2: Semantic Fidelity}) reports must faithfully reflect semantic modifications made in the workspace \cite{shipman1999formality}; and (\textbf{P3: Transparent Inference}) the framework must demonstrate interpretability—showing how the LLM analyzes semantic interactions and translates them into comprehensible feedback, enabling users to verify their inputs are properly incorporated.
These principles form the essential foundation for effective LLM-powered report refinement.
According to the three principles, we designed and implemented the following experiments.

In the case study, we employed \textit{gpt-4o-mini} model and utilized \textit{The Sign of Crescent} dataset \cite{hughes2003discovery}. This dataset consists of 41 fictional intelligence reports detailing three coordinated terrorist attack plots in three U.S. cities, with each plot involving at least four suspicious individuals. 

\subsection{Quantitative Evaluation}
To evaluate how semantic interactions affect results, we created 35 pairs of original and modified workspaces, each containing a 10-document plot. These pairs incorporated 13 combinations of semantic interaction types and granularities—including highlights, notes, and cluster reorganizations—as well as control cases without semantic interactions. Our experimental scenarios encompassed adding, removing, and modifying these elements, both individually and in combinations. The detailed distribution of these test pairs is available in the supplementary materials. After randomizing the sequence of pairs, we used an LLM to generate refined reports for comparative analysis.
\subsubsection{\textbf{P1}: Targeted Refinement}
In the generated reports, each cluster corresponds to one paragraph, with the first paragraph providing a summary and the last offering a conclusion, following the Bottom Line Up Front (BLUF) structure.
For evaluating Principle 1, we must determine whether refinements occur exclusively in targeted sections. Report refinement offers considerable flexibility, as modifications can manifest in various ways—an added highlight might result in text being appended to an existing sentence or generate an entirely new sentence.
Given the complexity of establishing sentence-level ground truth, we opted for paragraph-level evaluation. When a user modifies objects within a cluster, these changes should be reflected in the corresponding paragraph, as well as in the introductory summary and concluding paragraphs. Therefore, in this context, a correctly refined section equals one paragraph. We define $N_{TPP}$ as the number of correctly refined sections, $N_{PP}$ as the total number of refined sections, and $N_{TP}$ as the number of sections that should be refined. Based on these metrics, we calculate precision, recall, and F1-score for Principle 1 as follows:

\begin{equation} 
\begin{aligned}
Precision &= N_{TPP}/N_{PP} \\
Recall &= N_{TPP}/N_{TP} 
\end{aligned}
\end{equation}

We recommend using F1-score as the optimal metric for method comparison, as it balances competing concerns: Baseline achieves high recall but compromises precision by affecting all sections indiscriminately, while \appnamep{} delivers high precision but suffers from low recall, missing sections that require refinement. Based on this comprehensive evaluation criterion, \appname{} performs best with the highest F1-score.
\begin{table}[h]
    \centering
    \caption{Targeted Refinement (P1) comparison across methods}
    \small
    \renewcommand{\arraystretch}{1.1} 
    \begin{tabular}{c|ccc}
        \hline
        Methods & Baseline & \appnamep{} & \appname{}\\
        \hline
        Precision & 0.752 & 0.975 & 0.951\\
        Recall & 1 & 0.652 & 0.831\\
        F1-score & 0.858 & 0.782 & \textbf{0.887}\\
        \hline
    \end{tabular}
    \label{tab:p1}
\end{table}

\subsubsection{\textbf{P2}: Semantic Fidelity}
For P2, we evaluate whether refinements maintain relevant connections to semantic interactions. To measure interaction-relevant edits, we calculate sentence-level precision as the ratio of relevant edited sentences $N_{TPS}$ to total edited sentences $N_{S}$. We assess recall by measuring the percentage of semantic interactions reflected in the refinement, calculated as the ratio of realized interactions $N_{RSI}$ to total interactions $N_{SI}$. Importantly, semantic interactions are counted not by user operations but by the number of objects (highlights, notes, clusters) that differ between original and current workspaces.
To determine the relevance between interactions and edits, we extract key elements from different interaction types: entities and citations from highlights, entities from notes, and names and locations from clusters. These extracted elements serve as the basis for identifying relevant sentences in the refined text.

\begin{equation} 
\begin{aligned}
Precision &= N_{TPS}/N_S\\
Recall &= N_{RSI}/N_{SI}
\end{aligned}
\end{equation}

The results (Table~\ref{tab:p2}) demonstrate that \appname{} achieves the highest F1-score by balancing more precise edits than the baseline with higher interaction recall than \appnamep{}. This indicates that \appname{} not only ensures more edits remain relevant to the original content but also more effectively incorporates semantic interactions into the refined reports.
\begin{table}[h]
    \centering
    \caption{Semantic fidelity (P2) comparison across methods}
    \small
    \renewcommand{\arraystretch}{1.1} 
    \begin{tabular}{c|ccc}
        \hline
        Methods & Baseline & \appnamep{} & \appname{}\\
        \hline
        Precision & 0.348 & 0.582 & 0.558\\
        Recall & 0.694 & 0.526 & 0.684\\
        F1-score & 0.463 & 0.553 & \textbf{0.614}\\
        \hline
    \end{tabular}
    \label{tab:p2}
\end{table}

\subsection{Qualitative Analysis}
\subsubsection{\textbf{P3}: Transparent Inference}
\label{sec: p3}
We compared the refined reports generated by \appname{} and \appnamep{} from our previous experiment and subsequently deployed \appname{} for a continuous sensemaking process across the entire dataset. The results demonstrate that \appname{} satisfies P3 by revealing the following patterns:

\textbf{\appname{} explains the refinement.}
The projection from semantic interactions to refinement is nonlinear and flexible, offering numerous possibilities that may result in asymmetrical outcomes. Users' ability to understand these refinements is crucial for effectively interacting for report refinement. The LLM analysis process, detailed in the supplemental material, presents LLM-inferred human intent alongside corresponding report edits, making refinements through semantic interactions more transparent to users.

\textbf{LLM analysis provides critical contextual information.} The refined reports in \appnamep{} demonstrate that semantic interactions alone can introduce problematic content that appears reasonable superficially but contains substantive issues. We identified two distinct patterns: 1) isolated interactions lacking contextual information—such as highlights and cluster member removal—frequently lead to vague changes, and 2) LLMs incorporate ambiguous content from notes that require additional context for proper interpretation, such as unclear abbreviated references (e.g., "It is suspected that M is buying C-4 explosive from H") without sufficient identification of who is "M" and "C". However, \appname{} can present contextual information and provide reasoning for why include these facts.

\begin{figure}
    \centering
    \includegraphics[width=0.8\linewidth]{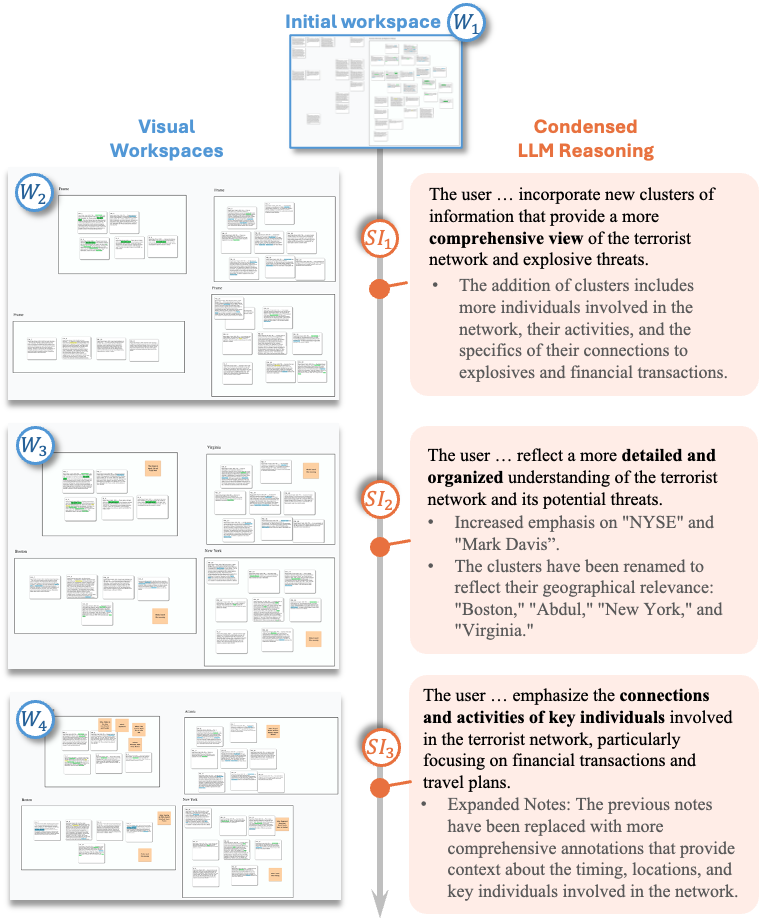}
    \caption{The condensed analysis, extracted from LLM analysis results, captures an iterative sensemaking timeline.}
    \label{fig:timeline}
\end{figure}

\textbf{Inference log timeline for sensemaking.}
Traditional interaction logs are difficult to review, while \appname{}'s LLM analysis provides intent summarization and analysis to address this limitation.
In Figure~\ref{fig:timeline}, we demonstrate an interactive sensemaking process using the full dataset, while presenting selected analytical outputs as a consolidated log.
The timeline enables users to review and reflect on each step of their sensemaking journey, allowing them to effectively track and understand their engagement patterns.

\section{Discussion}
\textbf{\appname{} Outperforms Other Methods.}
Both quantitative and qualitative analyses confirm \appname{}'s best performance across evaluation metrics. For diverse interaction types, \appname{} successfully automates the report refinement pipeline without manual prompting, provides transparent analytical processes, and delivers optimal results. The LLM analysis agent functions as the critical bridge connecting all essential components within the \appname{} framework. This approach presents a novel perspective for Human-AI collaboration systems, demonstrating the value of an agent capable of inferring and summarizing human intent to deliver tailored commands to other agents for sequential task execution.

\textbf{Semantic interactions in LLM-powered systems.} \appname{} obtained semantic interactions after comparing workspaces, enabling the LLM to make precise partial refinements. More significantly, leveraging semantic interactions enables LLM-powered systems to interpret users' sequential actions and provide contextually appropriate assistance. We explored alternative approaches, such as inputting pairs of workspaces to LLMs, but found that LLMs struggle to reliably identify differences compared to programmatic solutions. Since we only tested several semantic interactions in this study, transforming more interaction types will require further experimentation across diverse contexts. 

\textbf{Human Interaction Analysis.} Our results indicate that analyzing human interaction can significantly enhance report refinement. Beyond visual workspaces for sensemaking, this approach can be leveraged across various human-AI collaborative systems. It enables real-time analysis and expands possibilities for mixed-initiative systems that were previously rule-based. This method also facilitates understanding of user strategies, enabling personalized AI assistance. Additionally, it supports incremental formalism, while the inference log timeline helps users review and reflect on previous steps and verify alignment with goals. 
This approach's ability to infer intent extends its applications beyond sensemaking to decision making, problem solving, and other domains.
\textbf{Potential Risks.}
LLMs can provide seemingly reasonable inferences and refinements even with problematic workspaces, allowing incorrect user interactions to still generate plausible reports.
Balancing appropriate trust against over-reliance on AI content remains challenging. Verification mechanisms, transparent processes, and user validation could mitigate these risks.


\textbf{Limitations.}
Our method is not perfect. In the iterative refinement, we found the final refined report is messy with obvious text additions, while the one-time generated report is more concise and readable. The global or incremental generalization is one important question to consider in the future systems. The generation speed is another issue in which it will take about 20 seconds for the report refinement for a 10-document workspace generation. We may introduce more efficient methods like RAG \cite{lewis2020retrieval} in the future. Also, if the LLM inferred intent aligns with human real intent is unknown. Future work will address these limitations.

\section{Conclusion}
In this paper, we presented \appname{}, a framework that leveraging semantic interactions in visual workspaces for LLM report refinement, while addressing the challenge of inconsistent LLM-generated reports during refinement. Our case study demonstrated that this approach successfully better satisfies the three tested principles, handles various interaction types while providing transparent analysis and inference of human interactions to guide LLM refinement. 




\bibliographystyle{abbrv-doi}

\bibliography{main}

\begin{thebibliography}{10}

\bibitem{andrews2010space}
C.~Andrews, A.~Endert, and C.~North.
\newblock Space to think: large high-resolution displays for sensemaking.
\newblock In {\em Proceedings of the SIGCHI conference on human factors in computing systems}, pp. 55--64, 2010.

\bibitem{birnholtz2012tracking}
J.~Birnholtz and S.~Ibara.
\newblock Tracking changes in collaborative writing: edits, visibility and group maintenance.
\newblock In {\em Proceedings of the ACM 2012 conference on computer supported cooperative work}, pp. 809--818, 2012.

\bibitem{buschek2024collage}
D.~Buschek.
\newblock Collage is the new writing: Exploring the fragmentation of text and user interfaces in ai tools.
\newblock In {\em Proceedings of the 2024 ACM Designing Interactive Systems Conference}, pp. 2719--2737, 2024.

\bibitem{endert2012semantic}
A.~Endert, P.~Fiaux, and C.~North.
\newblock Semantic interaction for sensemaking: inferring analytical reasoning for model steering.
\newblock {\em IEEE Transactions on Visualization and Computer Graphics}, 18(12):2879--2888, 2012.

\bibitem{gero2024supporting}
K.~I. Gero, C.~Swoopes, Z.~Gu, J.~K. Kummerfeld, and E.~L. Glassman.
\newblock Supporting sensemaking of large language model outputs at scale.
\newblock In {\em Proceedings of the CHI Conference on Human Factors in Computing Systems}, pp. 1--21, 2024.

\bibitem{gupta2019abstractive}
S.~Gupta and S.~K. Gupta.
\newblock Abstractive summarization: An overview of the state of the art.
\newblock {\em Expert Systems with Applications}, 121:49--65, 2019.

\bibitem{hughes2003discovery}
F.~Hughes and D.~Schum.
\newblock Discovery-proof-choice, the art and science of the process of intelligence analysis-preparing for the future of intelligence analysis.
\newblock {\em Washington, DC: Joint Military Intelligence College}, 2003.

\bibitem{kang2023synergi}
H.~B. Kang, T.~Wu, J.~C. Chang, and A.~Kittur.
\newblock Synergi: A mixed-initiative system for scholarly synthesis and sensemaking.
\newblock In {\em Proceedings of the 36th Annual ACM Symposium on User Interface Software and Technology}, pp. 1--19, 2023.

\bibitem{lee2024design}
M.~Lee, K.~I. Gero, J.~J.~Y. Chung, S.~B. Shum, V.~Raheja, H.~Shen, S.~Venugopalan, T.~Wambsganss, D.~Zhou, E.~A. Alghamdi, et~al.
\newblock A design space for intelligent and interactive writing assistants.
\newblock In {\em Proceedings of the 2024 CHI Conference on Human Factors in Computing Systems}, pp. 1--35, 2024.

\bibitem{lewis2020retrieval}
P.~Lewis, E.~Perez, A.~Piktus, F.~Petroni, V.~Karpukhin, N.~Goyal, H.~K{\"u}ttler, M.~Lewis, W.-t. Yih, T.~Rockt{\"a}schel, et~al.
\newblock Retrieval-augmented generation for knowledge-intensive nlp tasks.
\newblock {\em Advances in neural information processing systems}, 33:9459--9474, 2020.

\bibitem{li2024superfiltering}
M.~Li, Y.~Zhang, S.~He, Z.~Li, H.~Zhao, J.~Wang, N.~Cheng, and T.~Zhou.
\newblock Superfiltering: Weak-to-strong data filtering for fast instruction-tuning.
\newblock {\em arXiv preprint arXiv:2402.00530}, 2024.

\bibitem{masson2024visual}
D.~Masson, Z.~Zhao, and F.~Chevalier.
\newblock Visual writing: Writing by manipulating visual representations of stories.
\newblock {\em arXiv preprint arXiv:2410.07486}, 2024.

\bibitem{pirolli2005sensemaking}
P.~Pirolli and S.~Card.
\newblock The sensemaking process and leverage points for analyst technology as identified through cognitive task analysis.
\newblock In {\em Proceedings of international conference on intelligence analysis}, vol.~5, pp. 2--4. McLean, VA, USA, 2005.

\bibitem{Santosh2025LexGenieAG}
T.~Y. S.~S. Santosh, M.~Aly, O.~Ichim, and M.~Grabmair.
\newblock Lexgenie: Automated generation of structured reports for european court of human rights case law.
\newblock 2025.

\bibitem{shipman1999formality}
F.~M. Shipman and C.~C. Marshall.
\newblock Formality considered harmful: Experiences, emerging themes, and directions on the use of formal representations in interactive systems.
\newblock {\em Computer Supported Cooperative Work (CSCW)}, 8:333--352, 1999.

\bibitem{suh2023sensecape}
S.~Suh, B.~Min, S.~Palani, and H.~Xia.
\newblock Sensecape: Enabling multilevel exploration and sensemaking with large language models.
\newblock {\em arXiv preprint arXiv:2305.11483}, 2023.

\bibitem{tang2024steering}
X.~Tang, E.~Krokos, C.~Liu, K.~Davidson, K.~Whitley, N.~Ramakrishnan, and C.~North.
\newblock Steering llm summarization with visual workspaces for sensemaking.
\newblock {\em arXiv preprint arXiv:2409.17289}, 2024.

\bibitem{tang2025respire}
X.~Tang, E.~Krokos, K.~Whitley, et~al.
\newblock Respire: Transparent and steerable human-ai sensemaking through shared workspace.
\newblock {\em TechRxiv}, April 2025. doi: {{%
10\hspace{.1pt}\discretionary{.}{%
}{.}\hspace{.4pt}36227\discretionary{/}{%
}{/}techrxiv\hspace{.1pt}\discretionary{.}{%
}{.}\hspace{.4pt}174438673\hspace{.1pt}\discretionary{.}{%
}{.}\hspace{.4pt}31381875\discretionary{/}{%
}{/}v1}}


\bibitem{yao2023react}
S.~Yao, J.~Zhao, D.~Yu, N.~Du, I.~Shafran, K.~Narasimhan, and Y.~Cao.
\newblock React: Synergizing reasoning and acting in language models.
\newblock In {\em International Conference on Learning Representations (ICLR)}, 2023.

\bibitem{zhang2023concepteva}
X.~Zhang, J.~Li, P.-W. Chi, S.~Chandrasegaran, and K.-L. Ma.
\newblock Concepteva: Concept-based interactive exploration and customization of document summaries.
\newblock In {\em Proceedings of the 2023 CHI Conference on Human Factors in Computing Systems}, pp. 1--16, 2023.

\bibitem{zhang2023visar}
Z.~Zhang, J.~Gao, R.~S. Dhaliwal, and T.~J.-J. Li.
\newblock Visar: A human-ai argumentative writing assistant with visual programming and rapid draft prototyping.
\newblock In {\em Proceedings of the 36th annual ACM symposium on user interface software and technology}, pp. 1--30, 2023.

\end{thebibliography}

\end{document}